\documentclass[14pt
reprint,%
aps,%
groupedaddress,%
superscriptaddress,%
twocolumn,%
assume,%
amsmath,%
floatfix%,
]{revtex4-2}

\usepackage{graphicx}
\usepackage[
  colorlinks=true,
  urlcolor=blue,
  linkcolor=blue,
  citecolor=blue
]{hyperref}
\usepackage{siunitx,braket,amssymb,gensymb,ulem,color}

\newcommand{\rrm}[1]{\textrm{#1}}

\newcommand{\lr}[1]{\left(#1\right)}

\newcommand{\dr}{\lr{\bm{r}}}

\newcommand{\lrr}[1]{\left[#1\right]}
\newcommand{\ul}{\bm{u}_1}
\newcommand{\uh}{\bm{u}_2}

\usepackage{graphicx}% Include figure files
\usepackage{dcolumn}% Align table columns on decimal point
\usepackage{bm}% bold math
\usepackage{ORCIDinREVTeX}

\begin{document}

\preprint{APS/123-QED}

\title{Unfolding unstable skyrmionic polarization textures}

%\title{\YS{Is skyrmion topology of optical polarization textures conserved?}}

\author{Nilo Mata-Cervera}
\orcid{0000-0001-8464-5102}
\email{nilo001@e.ntu.edu.sg}
\affiliation{Centre for Disruptive Photonic Technologies, School of Physical and Mathematical Sciences, Nanyang Technological University, Singapore 637371, Republic of Singapore}
\affiliation{Complex Systems Group, ETSIME, Universidad Politécnica de Madrid, Ríos Rosas 21, 28003 Madrid, Spain}
\author{Zhenyu Guo}
\orcid{0000-0003-4375-1560}
\email{zhenyu.guo@ntu.edu.sg}
\affiliation{Centre for Disruptive Photonic Technologies, School of Physical and Mathematical Sciences, Nanyang Technological University, Singapore 637371, Republic of Singapore}
\author{Yijie Shen}
\orcid{0000-0002-6700-9902}
\email{yijie.shen@ntu.edu.sg}
\affiliation{Centre for Disruptive Photonic Technologies, School of Physical and Mathematical Sciences, Nanyang Technological University, Singapore 637371, Republic of Singapore}
\affiliation{School of Electrical and Electronic Engineering, Nanyang Technological University, Singapore 639798, Republic of Singapore}

\begin{abstract}
%Stokes skyrmions are polarization textures of light which unwrap the Poincar\'e sphere in the transverse plane.
% \YS{Polarization textures of structured light can form skyrmion topology as that in condensed matter, however, the topological stability and conservation laws of which are in debate.}
Polarization of light can form skyrmionic textures, akin to nonlinear topological solitons in condensed matter, yet their disparate physical context has motivated extensive debate regarding their stability. Such optical polarization textures are usually constructed by superposing two spatial eigenmodes with different polarization. Here we show that the topological charge of such structures (skyrmion number) changes when an arbitrarily small perturbation splits coalescent phase singularities. In a superposition of two vortex beams, the skyrmion number generally only depends on the highest-order vortex topological charge $\lrr{Q_{\rm sk}=\max\lr{\ell_2,\ell_1}}$ rather than the difference of charges of the vortices in superposition $\lrr{Q_{\rm sk}=\ell_2-\ell_1}$, which only arises in an idealized case without perturbation. These results have significant implications for polarization structures with wavelength-scale localization and those experiencing complex aberrations.
\end{abstract}
\maketitle

% Topology underlies deeply in our physical world, where identical mathematical ingredients describe properties of a myriad of physical systems
Topology lies profoundly in our physical world, where identical mathematical ingredients describe properties of a myriad of physical systems~\cite{Singh2019IntroductionTopology,2024KnottedFields}. In wave physics, topology plays an important role describing the intricate singular structure of wave fields: isolated nodal points~\cite{Wang2022TopologicalLight,Vernon20233DFields,Dennis2001TopologicalFields,Dennis2025WaveOn}, phase and polarization singularity lines~\cite{Berry2000PhaseWaves,Berry2001PolarizationWaves,Dennis2004LocalTwirl,Angelsky2023PolarizationAspects,Oholleran2009TopologyDarkness,Dennis2010IsolatedKnots,Berry2001KnottedWaves,Sugic2021Particle-likeLight}, topological textures~\cite{Ye2024TheorySkyrmions,Wu2025PhotonicMonopoles,Lin2025Space-timeCrystals,Marco20254DLight,Shen2025Free-spaceTutorial}, to name a few. In contrast with singularities and defects, topological textures refer to continuous fields where a vector smoothly changes orientation in physical space in the absence of any discontinuity. Mimicking the vector structure of topological excitations in magnetic materials~\cite{Bogdanov2020PhysicalSkyrmions,Bernevig2022ProgressMaterials}, optical skyrmions are topologically non-trivial textures in light which garnered significant interest in the recent years~\cite{Shen2024OpticalLight}. Their vector field unwraps a spherical parametric space into a physical space, a mapping whose degree (skyrmion number) identifies topologically distinct field configurations~\cite{Braun2012TopologicalSolitons}.

Pioneered by a seminal work from Beckley et al.~\cite{Beckley2010FullBeams}, full Poincar\'e beams (now renowned as Stokes skyrmions) are a class of optical skyrmions which map the Poincar\'e sphere (PS) into the transverse plane~\cite{Ye2024TheorySkyrmions}. They attracted extensive research due to their rich topological features~\cite{Ye2024TheorySkyrmions,Gao2020ParaxialBeams} which are relatively robust in propagation through free-space and complex media~\cite{Wang2024TopologicalMedia,Wang2025TheTurbulence,Guo2026TopologicalTurbulence} as other topological indices in optics~\cite{Gbur2008VortexConservation,Cheng2009PropagationAtmosphere}, at least in comparison with mode purity~\cite{Anguita2008Turbulence-inducedLink}. Yet, emerging from unbounded linear wave fields these textures are not granted energy stability, and the magnitude of the field can possible vanish at isolated points unwrapping the topological configuration~\cite{Alonso2026CoveringFields}. 

\begin{figure}[t!]
    \centering
    \includegraphics[width=\linewidth]{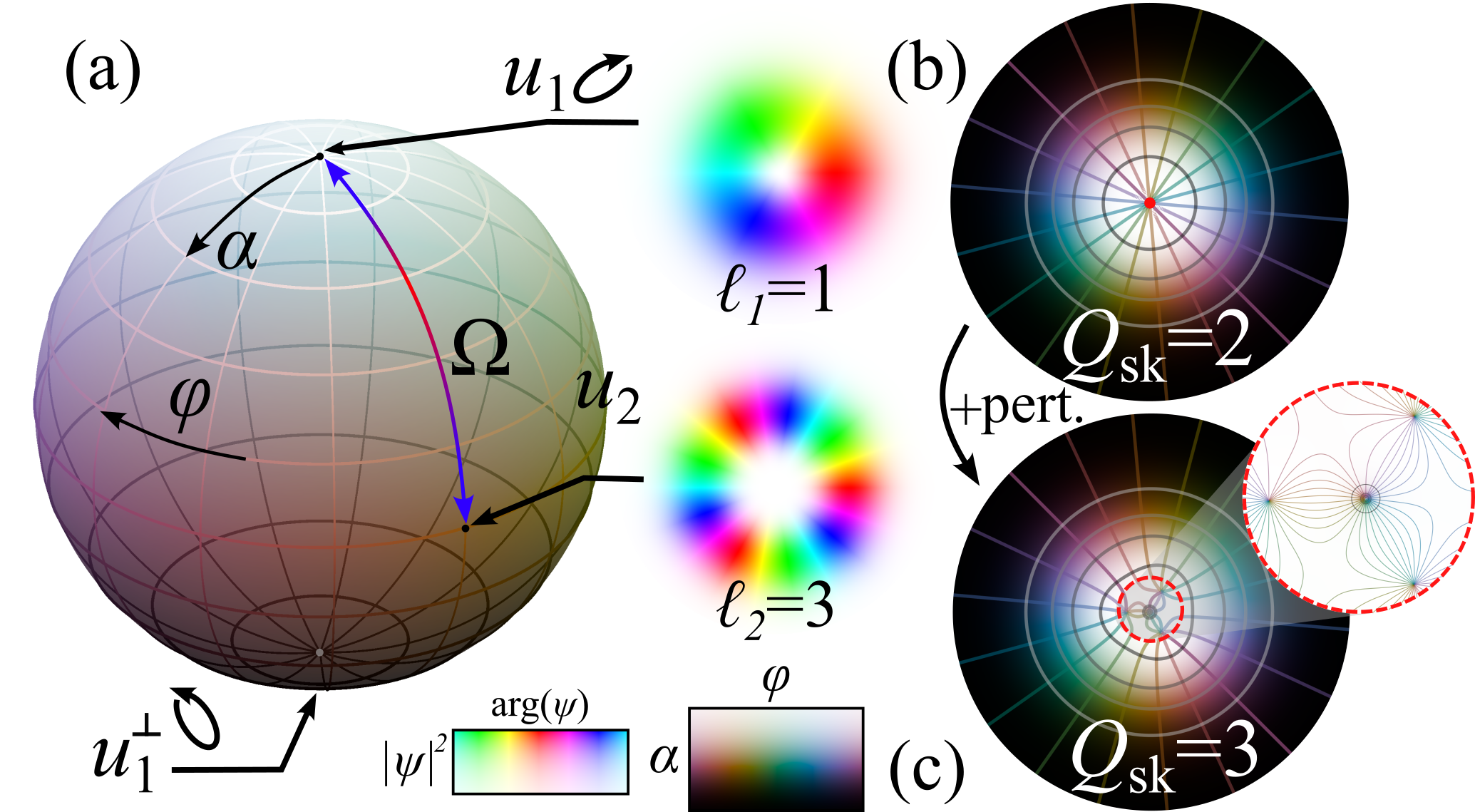}
    \caption{(a) Poincar\'e sphere (PS) in the rotated frame: $\ul$ and $\uh$ are the polarization states of $\psi_1$ and $\psi_2$ , with $\Omega$ the angle formed by their Stokes vectors, $\bm{u}_1^\perp$ the unit vector orthogonal to $\bm{u}_1$. $\alpha$ and $\varphi$ are the spherical angles. The coloured (gray) curves represent the meridians (parallels) of the PS. (b) Skyrmionic texture formed by $\ell_1=1$, $\ell_2=3$ without perturbation $\lr{\varepsilon=0}$ and $\Omega=\pi$ yielding $Q_{\rm sk}=\ell_2-\ell_1=2$. (c) Same with small perturbation $\lr{\varepsilon\neq0}$ where $Q_{\rm sk}=\rrm{max}\lr{\ell_1,\ell_2}=3$, with the additional polarization singularity (zero) highlighted as an inset.}
    % (d) Same as (c) with non-orthogonal polarizations $\Omega=\pi/2$.
    \label{F1}
\end{figure}

Here we show the unfolding of unstable skyrmionic polarization textures with an arbitrarily small perturbation splitting high-order phase singularities. The skyrmion number changes upon such perturbation, and is generally described only by the highest-order vortex topological charge $\lrr{Q_{\rm sk}=\max\lr{\ell_2,\ell_1}}$ rather than the difference of charges of the vortices in superposition $\lrr{Q_{\rm sk}=\ell_2-\ell_1}$, see Fig.~\ref{F1}. The latter only holds in the idealized case where high-order and low-order vortices of orthogonal polarizations coalesce, allowing convergence to a well-defined polarization state at a singularity of zero intensity [red dot in Fig.~\ref{F1}(b)]. However, this structure is not stable in a 2D plane~\cite{Spaegele2023TopologicallySpace,Vernon20233DFields,Nye1998UnfoldingDislocations}, unfolding upon infinitesimal perturbation into a stable configuration where the transverse field is nowhere zero. We illustrate our theoretical findings with numerical and experimental examples, and provide insights on why the widely used result $\lrr{Q_{\rm sk}\approx\ell_2-\ell_1}$ arises in many practical scenarios. 

Stokes skyrmions are usually described as a superposition of two vortex beams $\psi_1\dr$ and $\psi_2\dr$ with distinct topological charges $\ell_1$ and $\ell_2$ and polarizations $\ul$ and $\uh$~\cite{Gao2020ParaxialBeams}. Here we use the sub-indices ``1" and ``2" for the low-order and higher-order vortices ($|\ell_2|>|\ell_1|$), and focus on a transverse plane with coordinates $\bm{r}=\lr{x,y}$, $r=\sqrt{x^2+y^2}$, $\phi=\tan^{-1}\lr{y/x}$. Ignoring a global Gaussian envelope, a polynomial behaviour $[\psi_1\propto(r/w)^{|\ell_1|}\exp\lr{i\ell_1\phi}$ and $\psi_2\propto(r/w)^{|\ell_2|}\exp\lr{i\ell_2\phi}]$ guarantees that the relative weight of the two vortices $\psi_2/\psi_1$ covers all complex amplitudes ($\mathbb{C}_\infty$) when spanning the entire transverse plane ($\mathbb{R}^2_\infty$). Since each vortex represents the complex amplitude of the $\ul$ and $\uh$ polarization components, the latter implies all polarization states are monotonically covered in the transverse plane ($\mathbb{C}_\infty$ is the stereographic projection of the PS~\cite{Poincare1889TheoriePremier}). 
%The same ingredients can be applied to describe other topological polarization textures~\cite{Marco20254DLight,Mata-Cervera2025SkyrmionicVortices}.

% The lower-order vortex dominates over the higher-order vortex at the origin, while the higher-order vortex dominates over the lower-order vortex at infinity, formally $|\psi_2/\psi_1|\lr{r/w\ll1}\to0$ and $|\psi_2/\psi_1|\lr{r/w\gg1}\to\infty$.

We express the vector wave function as
\begin{equation}\label{eq:wave}
    \bm{\Psi}\dr=e^{i\theta}\sqrt{P}\psi_1\dr\ul+
\psi_2\dr\uh,
\end{equation}
where $\psi_1\dr$ and $\psi_2\dr$ are set as Laguerre-Gaussian (LG) vortices $\rrm{LG}_{\ell}^p\dr$ with azimuthal indices $\ell_1$ and $\ell_2$ and zero radial order, $p=0$. The Jones vectors $\ul$ and $\uh$ do not have to be orthogonal, therefore the angle $\Omega$ made by their Stokes vectors in the Poincar\'e sphere satisfies $0<\Omega\leq\pi$, see Fig.~\ref{F1}(a). $\sqrt{P}>0$ and $\theta$ are global amplitude and phase factors between the two vortices which we shall ignore. To evaluate the skyrmion number of the polarization texture in (\ref{eq:wave}) we first need to express the spatially-varying polarization state. We use the complex ratio of field amplitudes in an orthogonal polarization basis, for convenience $\ul$ and its orthogonal state $\ul^\perp$ (since skyrmionic properties do not depend on rotations of the PS we shall use any angular origin in the PS~\cite{Maxwell2025StochasticFields}). In general these two states are elliptical, and are antipodal at the PS since $\langle\ul|\ul^\perp\rangle=0$, see Fig.~\ref{F1}(a). Once $\ul$ and $\ul^\perp$ are known, their corresponding Stokes vectors define the spherical coordinate system in the PS: the polar angle origin $\alpha=0$ corresponds to $\ul$ and $\alpha=\pi$ corresponds to $\ul^\perp$. The azimuthal angle $\varphi$ is defined with respect to the Stokes vector of $\ul$, increasing in the clockwise sense, see Fig~\ref{F1}(a). 

All information of the spatially-varying polarization state is captured by the complex rational function~\cite{Poincare1889TheoriePremier} 
\begin{equation}\label{eq:complex_ratio}
    \tilde{f}\dr=\rho e^{i\gamma}=\frac{\langle\ul|\bm{\Psi}\rangle}{\langle\ul^\perp|\bm{\Psi}\rangle}=\frac{\sqrt{P}e^{i\theta}\psi_1+\tilde{\mu}\psi_2}{\tilde{g}\psi_2},
\end{equation}
whose modulus and argument are denoted by $\rho$ and $\gamma$. Here $\tilde{g}=\langle\ul^\perp|\uh\rangle$ and $\tilde{\mu}=\langle\ul|\uh\rangle$ are the (complex) projection coefficients of $\uh$ onto the orthogonal basis $\lr{\ul,\ul^\perp}$. Their amplitude is simply related to the angle $\Omega$ in the PS [Fig.~\ref{F1} (a)], namely $|\tilde{\mu}|=\cos\lr{\Omega/2}$ and $|\tilde{g}|=\sin\lr{\Omega/2}$~\cite{Alonso2023GeometricTutorial}. When the two vortices are shaped in orthogonal polarizations ($\Omega=\pi$ and $\tilde{\mu}=0$) the spherical coordinates in the sphere are simply given by the ratio between the two vortices' complex amplitude, but in general a cross-term $\tilde{\mu}\neq0$ appears (\ref{eq:complex_ratio}). The spherical angles measured with respect to the Stokes vector of $\ul$ are obtained from the inverse stereographic projection from (\ref{eq:complex_ratio}) to the unit sphere~\cite{Goldman1999ComplexGeometry}, 
\begin{equation}\label{eq:coordinates}
    \cos\lr{\alpha}=\frac{\rho^2-1}{\rho^2+1}, \hspace{0.75cm} \varphi=-\gamma.
\end{equation}
The zeros and poles of $\tilde{f}$ are polarization singularities in a generalized sense (points of undefined $\varphi$)~\cite{McWilliam2023TopologicalMulti-skyrmions}.

The polarization field (\ref{eq:complex_ratio}) has non-zero skyrmion number, which takes the form of a solid angle integral
\begin{equation}\label{eq:Q_sk_surface}
    Q_{\rm sk}=\frac{1}{4\pi}\int_{\sigma}\sin\alpha\lr{\bm{r}}\left|J\right|\rrm{d}x \rrm{d}y,
\end{equation}
with $J=\partial (\alpha,\varphi)/\partial(x,y)$ the Jacobian matrix with determinant $\left|J\right|=\partial_x\alpha\lr{\bm{r}}\partial_y\varphi\lr{\bm{r}}-\partial_y\alpha\lr{\bm{r}}\partial_x\varphi\lr{\bm{r}}$, determining the rate of coverage of the PS per unit area in the plane. The skyrmion number (topological degree) is an integer counting the number of times the PS is wrapped over a spatial sphere, with $\sigma$ the compact region of the transverse plane which can be mapped to a sphere. For simplicity, we use the same symbol $Q_{\rm sk}$ for the skyrmion number of a polarization texture spanning the extended transverse plane (with uniform polarization at infinity) and for the degree of coverage of the PS in a finite-size polarization texture (with varying polarization state at the boundary). Although full Poincar\'e beams were initially constructed using a vortex beam and a Gaussian beam~\cite{Beckley2010FullBeams}, similar topologies have been proposed with both beams having non-zero vortex charge~\cite{Wang2025GenerationNumber}. Such construction is peculiar since the point $r=0$ in (\ref{eq:wave}) carries no intensity, where both polarization components vanish simultaneously. Only the idealization of high-order vortex cores allows eluding a discontinuity at $r=0$ since the polarization state converges smoothly around this point. Integrating (\ref{eq:Q_sk_surface}) over the entire transverse plane yields a skyrmion number $Q_{\rm sk}=\ell_2-\ell_1$ for the beam superposition (\ref{eq:wave}), as can be widely found in the literature~\cite{Gao2020ParaxialBeams,Ye2024TheorySkyrmions,Shen2025Free-spaceTutorial}. 

% To accommodate the singular point at $r=0$ where the light intensity vanishes and the polarization is not defined, $\sigma^2$ represents a compact disk in the transverse plane which avoids $r=0$. Such quotient space $\sigma^2/\mathcal{S}^1$ (where $\mathcal{S}^1=\partial\sigma^2$) is homeomorphic to $\mathbb{S}^2$~\cite{Singh2019IntroductionTopology}. Eq. (\ref{eq:Q_sk_surface}) can be integrated analytically over the entire disk $\sigma^2$ yielding $Q_{\rm sk}=\ell_2-\ell_1$, as can be widely found in the literature~\cite{Gao2020ParaxialBeams,Ye2024TheorySkyrmions,Shen2025Free-spaceTutorial}. 
% \red{We must correct this. The punctured transverse plane is not equivalent to the extended transverse plane. We must say that at the singularity $r\to0$ we remove the singularity as $\bm{S}(0)=\lim_{r\to0}\bm{S}\dr$ and thus we work with the extended transverse plane and not the punctured extended transverse plane.}

Although the presence of a singularity with zero intensity is problematic, this is only an artifact from the idealization of (\ref{eq:wave}) which in a more general case has a richer topological structure. In fact, the event where zeros of both transverse fields in (\ref{eq:wave}) fall at the same position breaks down upon infinitesimal perturbation. Even in a lower-order case, a cylindrical vector beam, the coalescence of two first-order zeros of orthogonal polarization states have codimension 4: two real conditions on each polarization component must be satisfied simultaneously for the intensity to vanish at a point, a singularity that is not stable in a 2D plane~\cite{Dennis2001TopologicalFields}. A more general situation is to consider Eq. (\ref{eq:wave}) perturbed by an arbitrarily weak wave $\tilde{\bm{\varepsilon}}=\tilde{\varepsilon}_1\ul+\tilde{\varepsilon}_2\uh$ which can account for the splitting of unstable high-order phase singularities~\cite{Nye1998UnfoldingDislocations}. The arbitrariness of $\bm{\varepsilon}$ implies that the nodes of $\psi_1+\tilde{\varepsilon}_1$ and $\psi_2+\tilde{\varepsilon}_2$ would generally fall at different positions of the transverse plane, otherwise $\tilde{\bm{\varepsilon}}$ would need to be finely tuned. It follows that splitting the singularities of (\ref{eq:wave}) does not preserve the skyrmion number (\ref{eq:Q_sk_surface}), as conceptually shown in Fig.~\ref{F1}(b-c). 

Following the approach by McWilliam et al~\cite{McWilliam2023TopologicalMulti-skyrmions}, we write Eq. (\ref{eq:Q_sk_surface}) as a line integral along a contour $\Gamma=\partial\sigma$ which encloses the region $\sigma$ and surrounds all polarization singularities. Using our notation the line integral is given by 
$$Q_{\rm sk}=\frac{1}{4\pi}\oint_\Gamma \cos{\alpha}\dr\nabla\varphi\dr\rrm{d}\bm{r}=$$
\begin{equation}\label{eq:Q_sk_line}
    =\frac{1}{2}\lr{\sum_j \cos{\alpha}^{(j)}q^{(j)}-\cos{\alpha}^{(\infty)}q^{(\infty)}},
\end{equation}
provided that the polarization state converges to a constant value $\cos\alpha^{\lr{\infty}}$ at the boundary $\Gamma$. The integers $q^{(j)}=(2\pi)^{-1}\oint_j\nabla\varphi\dr d\bm{r}=-(2\pi)^{-1}\oint_j\nabla\gamma\dr d\bm{r}$ are simply the topological charges of $\tilde{f}$ around each of its $j$ phase singularities (both zeros and poles). The polar angle reads $\alpha^{(j)}=0$ $\lr{\alpha^{(j)}=\pi}$ at the poles (zeros) of $\tilde{f}$, which further simplifies (\ref{eq:Q_sk_line}) into $Q_{\rm sk}=(1/2)\lr{N_p-N_z-\cos{\alpha}^{(\infty)}q^{(\infty)}}$. The key aspect here is to treat zeros and poles independently, in other words, the nodes of the numerator and denominator in (\ref{eq:complex_ratio}) do not coalesce, which agrees with the codimension argument stated before. Accordingly the path $\Gamma$ encloses the entire area of $\sigma$ where there are no singular points of undefined polarization state, only polarization singularities of undefined $\varphi$.

Since $\tilde{f}$ is meromorphic, its winding number is given by Cauchy's argument principle, 
\begin{equation}\label{eq:Cauchy}
    \frac{1}{2\pi i}\oint_{\Gamma} \frac{\tilde{f}'\dr}{\tilde{f}\dr}d\bm{l}=\frac{1}{2\pi}\oint_{\Gamma}\nabla\gamma\dr d\bm{l}
=N_z+N_p,
\end{equation}
with $N_z$ and $N_p$ again the zeros and poles of the function inside the domain delimited by $\Gamma$. 
% Note that we write the net topological charge as the sum of poles and zeros of $\tilde{f}$ inside $\Gamma$, and not the difference between poles and zeros common in complex analysis.
\begin{figure}[t!]
    \centering
    \includegraphics[width=\linewidth]{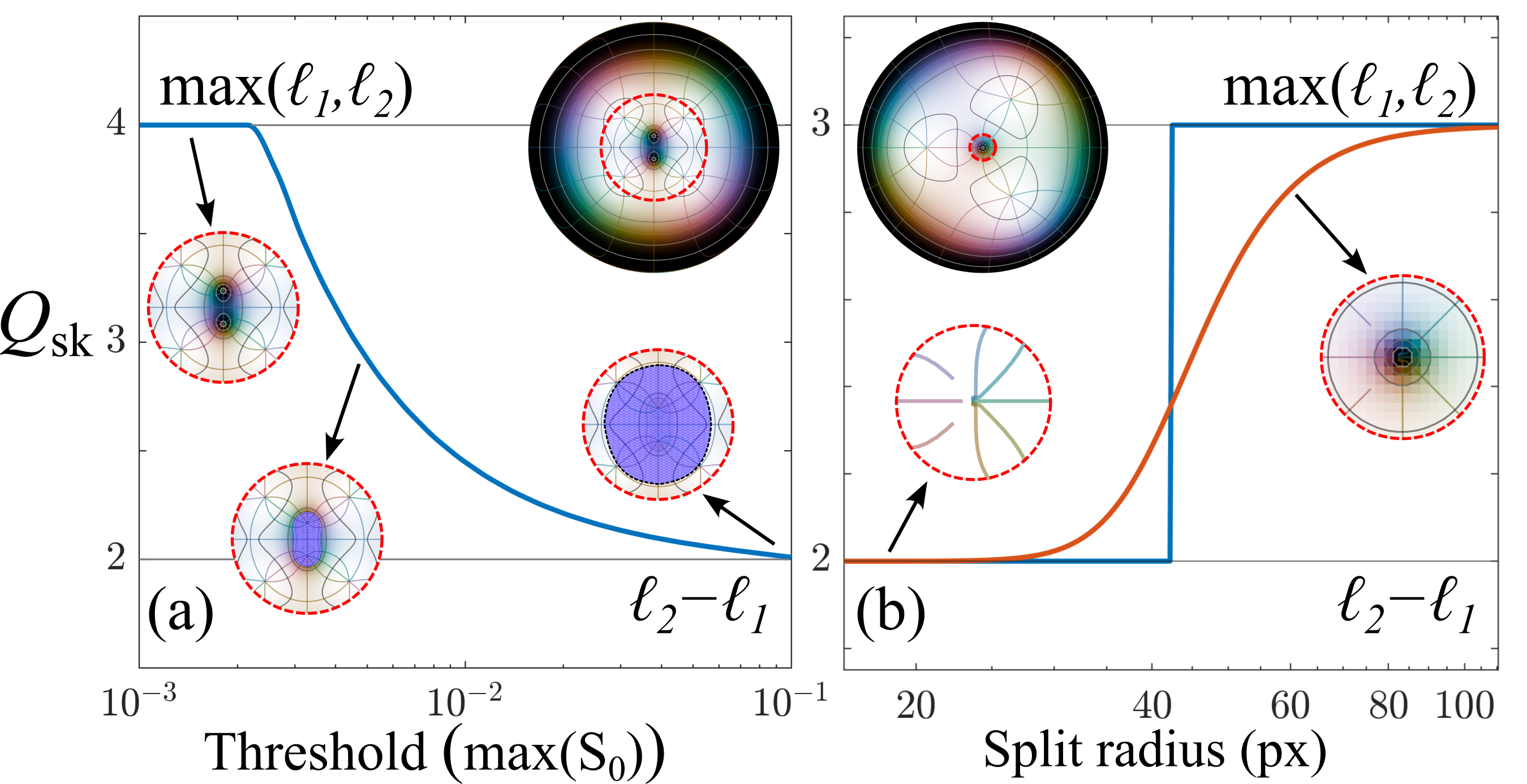}
    \caption{Numerically integrated PS coverage $Q_{\rm sk}$: (a) for increasing intensity threshold normalized to the maximum intensity ($\max\lrr{S_0}$), (b) for increasing split distance (in pixels) between singularities. In (b), we show the comparison between integration of Eq. (\ref{eq:Q_sk_surface}) with finite differences (red curve) and with plaquette method (blue curve). Eq. (\ref{eq:Q_sk_surface}) converges to the theoretical integer value $Q_{\rm sk}=\max\lr{\ell_2,\ell_1}$ unless regions of low intensity are cropped (a) or undersampled (b). Data (a): $\lr{\ell_1,\ell_2}=\lr{2,4}$, $\lr{p_1,p_2}=\lr{1,0}$, $\lr{\varepsilon_1,\varepsilon_2}=\lr{10^{-3},5\cdot10^{-4}}$. Data (b): $\lr{\ell_1,\ell_2}=\lr{1,3}$, $\lr{p_1,p_2}=\lr{1,0}$, $\varepsilon_1=0$, $\varepsilon_2\in\lrr{10^{-7},10^{-2}}$. In all simulation $\bm{u}_1=\bm{u}_R$ and $\bm{u}_2=\bm{u}_L$.}
    \label{F2}
\end{figure}
We can therefore invoke (\ref{eq:Cauchy}) to calculate $q^{(\infty)}$ in (\ref{eq:Q_sk_line}) obtaining 
\begin{equation}\label{eq:Qsk_Nzp}
    Q_{\rm sk}=\frac{1}{2}\lr{N_p-N_z-\cos{\alpha}^{(\infty)}\lr{N_z+N_p}}.
\end{equation}
Besides its simplicity, Eq. (\ref{eq:Qsk_Nzp}) has important consequences, since it implies that removing points of zero total intensity in (\ref{eq:wave}) does not preserve the skyrmion number. In our notation $\lr{|\ell_2|>|\ell_1|}$ and with orthogonal polarizations ($\tilde{\mu}=0$), the asymptotic dominance away from the origin implies $\cos{\alpha}^{(\infty)}=-1$ which simplifies (\ref{eq:Qsk_Nzp}) into our final expression 
\begin{equation}\label{eq:Qskmax}
    Q_{\rm sk}=N_p=\max\lr{\ell_1,\ell_2}=\ell_2.
\end{equation}
In other words, the skyrmion number only depends on the topological charge of the higher order vortex. Note that by $\max\lr{\bm{x}}$ we mean the element $x_i$ of $\bm{x}$ with maximum absolute value. This result is independent of whether the polarization states are orthogonal or not. Consider $\tilde{\mu}\neq0$ in (\ref{eq:complex_ratio}), which at infinity it implies $\tilde{f}=\tilde{\mu}/\tilde{g}$: there are no polarization singularities at infinity where the state is $\uh$ ($\alpha^{(\infty)}=\Omega$). Therefore $q^{(\infty)}=0\rightarrow N_z=-N_p$ and Eq. (\ref{eq:Qsk_Nzp}) yields again $Q_{\rm sk}=N_p=\ell_2$. This confirms that the skyrmion number only depends on the highest-order topological charge regardless of the degree of orthogonality of the polarization states $\ul$ and $\uh$, and vanishes if the polarization texture is trivial $\lr{\Omega=0}$. It is then straightforward to generalize the definition of skyrmionic beam to a more complex vortex superposition
\begin{equation}\label{eq:superp_sk}
    \psi_1\dr=\sum_{\ell=0}^{\ell_1}c_{\ell}\rrm{LG}^0_{\ell}\dr, \hspace{0.5cm} \psi_2\dr=\sum_{\ell=0}^{\ell_2}c_{\ell}\rrm{LG}^0_{\ell}\dr
\end{equation}
with $|\ell_2|>|\ell_1|$ and $c_{\ell}$ non-zero coefficients. When the coefficients $c_\ell$ are arbitrary, by the fundamental theorem of algebra the polynomials in (\ref{eq:superp_sk}) respectively have $\ell_1$ and $\ell_2$ first-order zeros. In such case, Eq. (\ref{eq:Qsk_Nzp}) yields again $Q_{\rm sk}=\rrm{max}\lr{\ell_2,\ell_1}$ thus demonstrating the generality of the result. Eq. (\ref{eq:Qskmax}) remains valid as long as the polarization state of the superposition (\ref{eq:superp_sk}) is well-defined everywhere. In the singular case where two vortices of orthogonal polarization states coalesce the polarization state is undefined and so the definition of the skyrmion number in Eq. (\ref{eq:Q_sk_surface}) is problematic, but this is not a generic event for monochromatic light in 3D space.

Fig.~\ref{F1} illustrates this with a simple numerical example: two orthogonally polarized $(\Omega=\pi)$ vortices $\psi_1$ and $\psi_2$ with $\lr{\ell_1,\ell_2}=\lr{2,3}$. In the unperturbed case (\ref{eq:Q_sk_surface}) yields $Q_{\rm sk}=\ell_2-\ell_1=2$ when integrated over a disk avoiding $r=0$ (b). When the vortices are minimally perturbed and the singularities split the skyrmion number jumps to the general value given by Eq. (\ref{eq:Qskmax}): $Q_{\rm sk}=\ell_2=3$ (c). Although the splitting changes the skyrmion number, it still preserves the Poincar\'e index of the field of polarization contours: the degenerate singularity (b) with index +1 unfolds (c) into 4 singularities and 3 saddles (inset), yielding overall index of $4-3=1$ identically to (b) as expected~\cite{Dennis2006RowsBeam,Freund1995SaddlesFields}. 
\begin{figure}[b!]
    \centering
    \includegraphics[width=\linewidth]{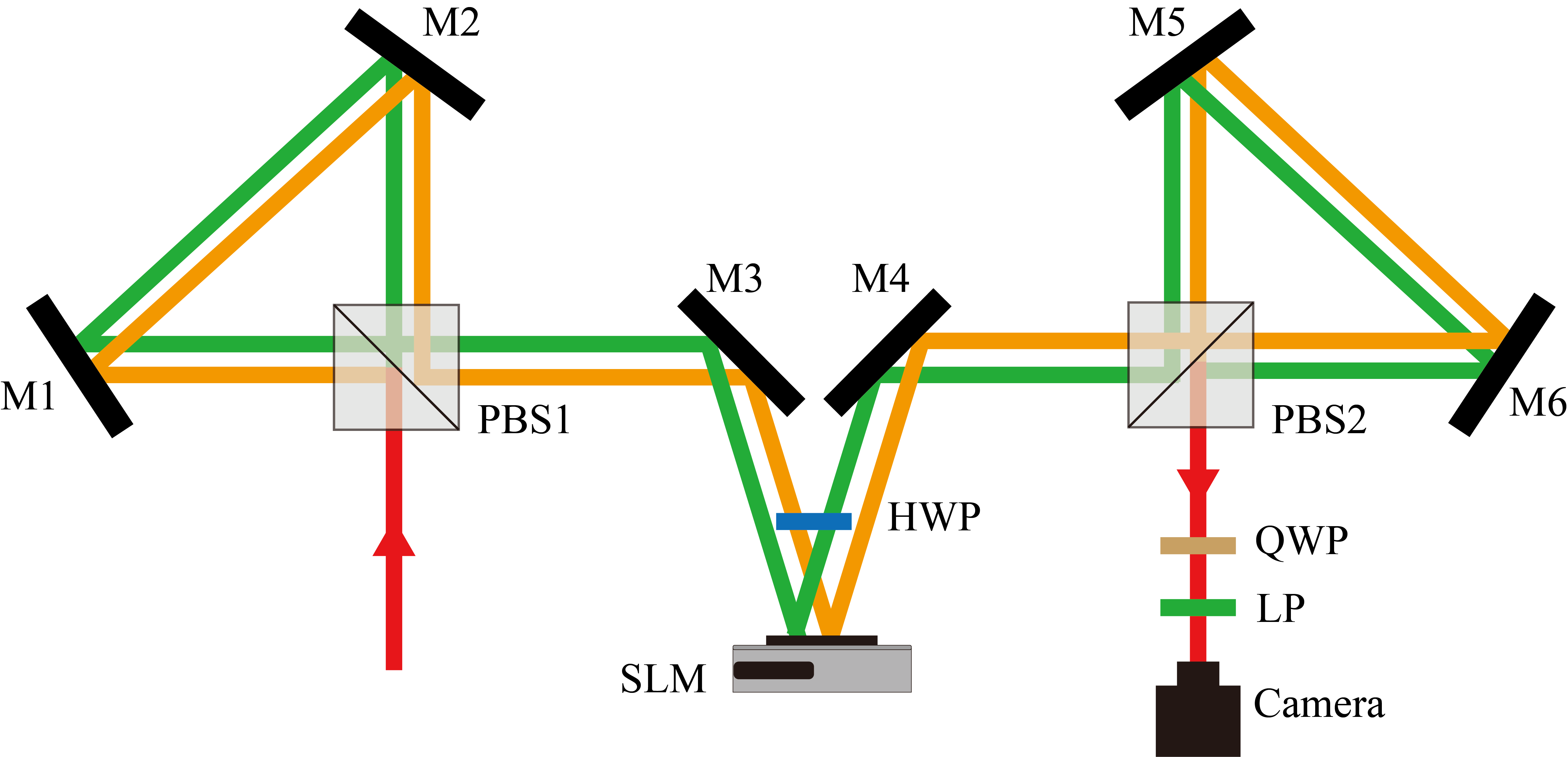}
    \caption{Experimental setup. PBS: polarizing beam splitter, M: mirror, HWP: half-wave plate, QWP: quarter-wave plate, LP: linear polarizer.}
    \label{F3}
\end{figure}

\begin{figure}[t!]
    \centering
    \includegraphics[width=\linewidth]{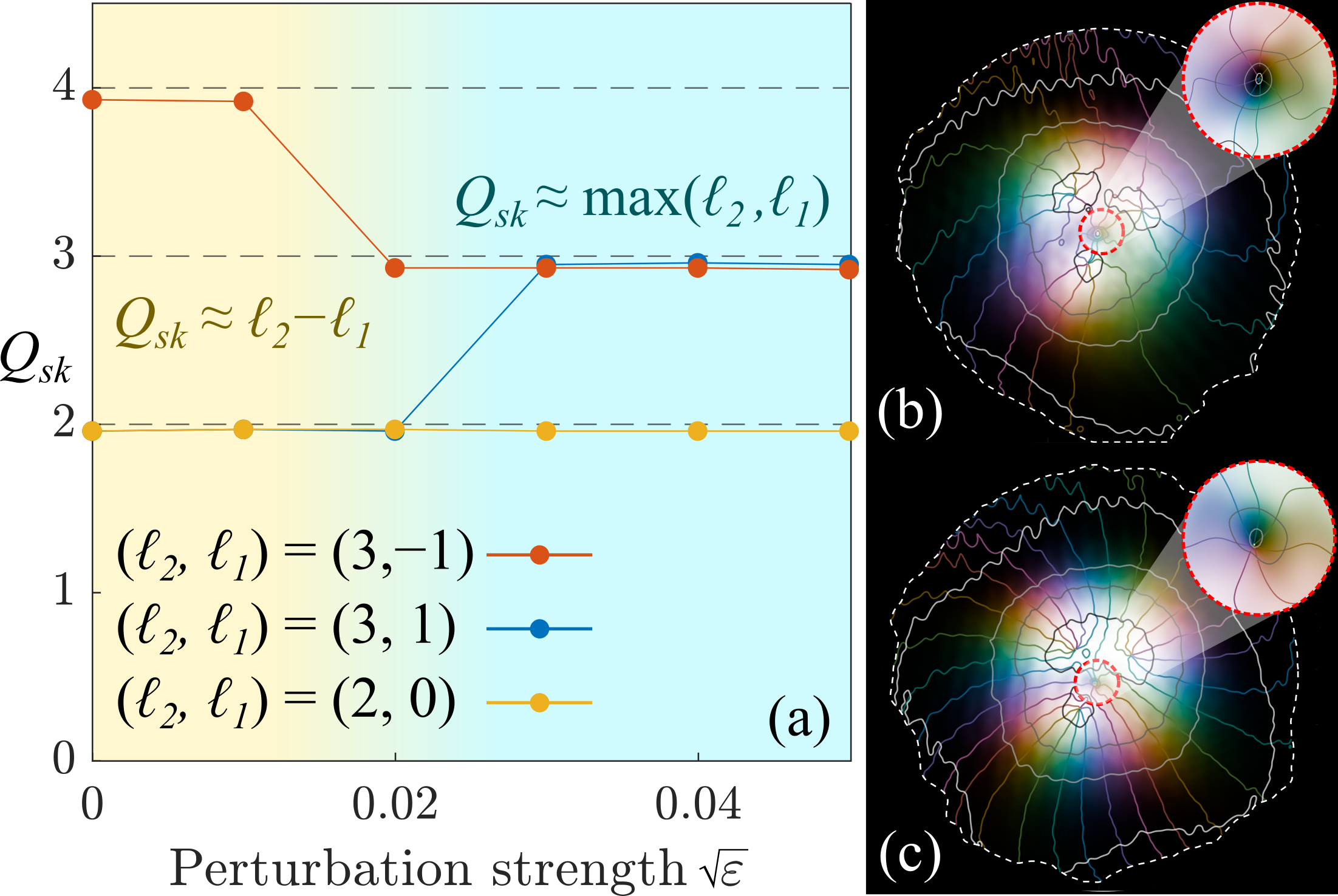}
    \caption{Experimentally retrieved PS coverage $Q_{\rm sk}$ for increasing perturbation strength $\sqrt{\varepsilon}=0,0.01,\dots,0.04, 0.05$ and different vortex superpositions (a). Polarization textures are shown for $\lr{\ell_1,\ell_2}=\lr{1,3}$ and $\sqrt{\varepsilon}=0.03$ (b) and for $\lr{\ell_1,\ell_2}=\lr{-1,3}$ and $\sqrt{\varepsilon}=0.02$ (c).} 
    % Top rows represent $\lr{\ell_1,\ell_2}=\lr{-1,3}$ and lower rows $\lr{\ell_1,\ell_2}=\lr{1,3}$. Numerically integrated PS coverage jumps from $Q_{\rm sk}\approx\ell_2-\ell_1$ to $Q_{\rm sk}\approx\max\lr{\ell_1,\ell_2}$ (up to experimental errors) when the additional polarization singularity is resolved by the detector (see inset). White dotted dashed contours delimit the region of interest.}
    \label{F4}
\end{figure}
Besides its generality, Eq. (\ref{eq:Qskmax}) can be circumvented due to two common situations. First, performing intensity thresholding of (\ref{eq:wave}), where the regions of low intensity are excluded in numerical integration, and second, when the separation between singularities is too small to be resolved. The former directly ignores the variations of polarization at regions of low intensity, while the latter is simply a sampling artifact, where the fast variations of polarization state near low intensity areas --- large values of the Jacobian $|J|$~\cite{Maxwell2025StochasticFields} --- are not captured by a typically low resolution mesh. 

We show these effects in a numerical simulation, the results are depicted in Fig.~\ref{F2}. We use two LG vortices, $\psi_1=\rrm{LG}_{\ell_1}^1\dr$ and $\psi_2=\rrm{LG}_{\ell_2}^0\dr$, with $\psi_1$ having non-zero radial index so that integration is limited to a disk of radius $w_0\sqrt{1+|\ell_1|}/\sqrt{2}$ with boundary conditions ($\psi_1=0$) at the disk edge, ensuring the integrality of $Q_{\rm sk}$~\cite{Wang2024TopologicalMedia}. Perturbation is introduced by a weak Gaussian beam $\tilde{\varepsilon}_{1,2}=\sqrt{\varepsilon_{1,2}}\ \rrm{LG}_0^0$, whose amplitude in the caption on Fig.~\ref{F2} is chosen arbitrarily just for illustration of the effect. The singularities split to a radius $r_{1,2}\propto \varepsilon_{1,2}^{|\ell_{1,2}|/2}$~\cite{Berry2001KnottedWaves}, shown normalized to the pixel size in (b). The modal cross-talk experienced by each vortex beam $P^{\rm cross}_{1,2}=|\langle\psi_{1,2}+\tilde{\varepsilon}_{1,2}|\tilde{\varepsilon}_{1,2}\rangle|^2$ is just $P^{\rm cross}_{1,2}=\varepsilon_{1,2}$, which is extremely low $\lr{<1\%}$ in all simulations.

In (a) the regions of low intensity (inset blue striped regions) are excluded during numerical integration, with the threshold normalized to $\rrm{max}\lr{S_0}$, with $S_0=|\bm{\Psi}|^2$ the total intensity of light. When the entire disk is considered, the integral (\ref{eq:Q_sk_surface}) yields $Q_{\rm sk}=\ell_2$, and gradually decays to $Q_{\rm sk}\approx\ell_2-\ell_1$ for increasing intensity threshold. Strictly speaking, intensity threshold itself neither guarantees integrality nor convergence to $Q_{\rm sk}\approx\ell_2-\ell_1$, and the inner boundary of varying polarization state should be found numerically choosing a contour of maximum $\cos{\alpha}$. Similarly, if the distance between the split phase singularities decreases to a single grid pixel level, numerical integration of (\ref{eq:Q_sk_surface}) may not capture the fast variations of polarization state and the (undersampled) integration of (\ref{eq:Q_sk_surface}) yields again $Q_{\rm sk}\approx\ell_2-\ell_1$ (b). In the case of a coarse spatial grid, the choice of an appropriate numerical method for integration of (\ref{eq:Q_sk_surface}) is crucial. As shown in (b), finite difference integration (red curve) is more susceptible to the grid resolution than the plaquette method (blue curve) described in~\cite{Berg1981Definition-model}. The core of the polarization textures with intensity threshold and undersampling are shown as insets.

We have experimentally verified the above using a Sagnac interferometer with a spatial light modulator (SLM) which generates arbitrary polarization textures~\cite{Gutierrez-Vega2017On-demandBeams}, the setup is shown in Fig.~\ref{F3}. The first Sagnac separates the incident linearly polarized beam ($\lambda=\SI{780}{\nano\metre}$) into two counter-propagating orthogonally polarized components. A half-wave plate aligns the polarization with the SLM (PLUTO-2.1-NIR-113) modulation axis, where the hologram independently controls the amplitude and phase of each component. After coherent recombination in the second Sagnac, the output beam has on-demand polarization texture. The Stokes parameters are reconstructed using a quarter-wave plate, linear polarizer, and a 12-bit CMOS camera from the intensity profiles in 6 polarization basis~\cite{Born1980PrinciplesLight}.

The results of the polarimetry measurements are shown in Fig.~\ref{F4} for different vortex superpositions and perturbation strengths. 
% where for simplicity we chose $\ul=\bm{u}_x$ and $\uh=\bm{u}_y$, $\lr{\ell_1,\ell_2}=\lr{-1,3}$ in (a) and $\lr{\ell_1,\ell_2}=\lr{+1,3}$ in (b). 
Both vortices are perturbed with a Gaussian beam with varying strength from $\sqrt{\varepsilon}=0$ to $\sqrt{\varepsilon}=0.05$. The PS coverage $Q_{\rm sk}$ shown in (a) is obtained by integrating (\ref{eq:Q_sk_surface}) in the region of interest, with its boundary calculated using an active contour model that minimizes $\cos\alpha^{(\infty)}$ along the contour~\cite{Kass1988Snakes:Models,Peters2026ExtractingTutorial}. We observe an abrupt jump of the numerically integrated PS coverage from $Q_{\rm sk}\approx\ell_2-\ell_1$ to $Q_{\rm sk}\approx\ell_2$ (up to numerical and experimental errors) once the perturbation strength exceeds a certain level (a). The abrupt jump is directly associated with the appearance of an additional zero of $\tilde{f}$, the new polarization singularity being highlighted as an inset in (b, c). Both the dynamic range and spatial resolution of the camera affect when this additional polarization singularity can be experimentally resolved, as it appears at regions where light intensity is low and the change of polarization state is very rapid. We also confirmed that if the splitting is experimentally resolved, further increase of the perturbation strength does not change $Q_{\rm sk}$, as seen in the blue and red curves in (a). Of course, if one of the beams has zero vortex charge $\lr{\ell_1=0}$ then the skyrmion number does not change $Q_{\rm sk}=\ell_2-\ell_1=\max\lr{\ell_2,\ell_1}=\ell_2$ with perturbation, as the yellow curve in (b) confirms. During post-processing, a Gaussian filter with a 5-pixel standard deviation smoothing kernel has been applied to all intensity images, and Eq. (\ref{eq:Q_sk_surface}) is integrated in a mesh with twice the resolution of the camera using spline interpolation. This is to reduce effects of background and quantization noise and mesh discretization in numerical calculations.

In conclusion, we have shown the unfolding of skyrmionic polarization textures when an arbitrarily small perturbation splits their unstable high-order polarization singularities. We demonstrate that the skyrmion number of a superposition of two vortices with distinct topological charges and polarizations is generally determined by the higher-order topological charge $\lrr{Q_{\rm sk}=\max\lr{\ell_1,\ell_2}}$ rather than the difference of charges $\lrr{Q_{\rm sk}=\ell_2-\ell_1}$ which strictly speaking only holds in the idealized case where the polarization field is ill-defined at the origin. This result can be generalized to superpositions of multiple vortex beams up to a finite azimuthal order (\ref{eq:superp_sk}), where the skyrmion number is invariant to the superposition coefficients. We also illustrated how the result $Q_{\rm sk}=\max\lr{\ell_1,\ell_2}$ arises when neglecting the fast variations of polarization state at regions of low intensity, or when these are undersampled by a low resolution numerical mesh. Rather than a theoretical entelechy, these results are a vivid manifestation of the rich characteristics of singular polarization fields that arise when minimally departing from ideal configurations. This simple example opens questions on the actual topology of skyrmionic textures after propagation through complex media. 

On the other hand, the topological transformation induced by perturbation of high-order vortices has a strong significance in skyrmionic textures based on three-dimensional spin angular momentum and those confined at wavelength scales, where the split of singularities cannot be neglected. As has been reported in several examples before, the skyrmion number of highly symmetric spin fields is bounded to unity~\cite{Mata-Cervera2025SkyrmionicVortices,Annenkova2025UniversalSound,Wu2025PhotonicMonopoles,Wang2022TopologicalLight}, yet their unfolding upon perturbation demonstrated here suggests that these structures are in practice not topologically limited.

% Several arguments can be made. First, new polarization singularities can be introduced in the field. These are not strange phenomena, since vortices are natural phenomena due to interference of waves. 

% Boundary conditions $(\cos\alpha^{(\infty)}=\pm1)$ are essential to guarantee integer coverages of the PS. However, in a free-space skyrmionic texture there is no physical law fixing the polarization state at the boundary. Only theoretically one of the polarization components has asymptotic dominance over the the other: away from the origin both $|\psi_1|$ and $|\psi_2|$ decay to zero but one of them does so at a faster rate than the other. In practice, this asymptotic dominance can be deteriorated or even reversed upon perturbation, so that strict integer wrappings are not protected in real settings. Therefore it is desirable to design (\ref{eq:wave}) so that the polarization ratio (\ref{eq:complex_ratio}) diverges to infinity as fast as possible therefore the convergence of (\ref{eq:Q_sk_line}) to integer values happens at regions of high intensity around the beam axis. The factors $\sqrt{P}$ and $\tilde{g}$ in (\ref{eq:complex_ratio}) play a crucial role in this regard, this is a direction we are currently exploring. The net zero topological charge of $\tilde{f}$ along the boundary implies that the lines of constant polarization azimuth do not meet at infinity, but instead they form closed loops [see Fig.~\ref{F1}(c)]. 

\noindent \textbf{Acknowledgements.} The authors acknowledge no conflict of interest. N. M.-C. acknowledges fruitful discussions with Shiqi Chen who drew our attention to Berg and L\"{u}scher paper~\cite{Berg1981Definition-model}.\\
\noindent \textbf{Author contributions.} N. M-C. conceived the idea, derived the theory, carried out numerical simulations, processed experimental data and wrote the paper. Z. G. prepared the experimental setup and carried out all measurements. Y. S. supervised the research. All authors participated in the discussion and analysis of the results.\\
\textbf{Funding.} Singapore Ministry of Education (MOE) AcRF Tier 1 grants (RG157/23 \& RT11/23), Singapore Agency for Science, Technology and Research (A*STAR) (M24N7c0080 \& H25-MRO3489), and Nanyang Assistant Professorship Start Up grant. Spanish Ministry of Science, Innovation and Universities (PID2025-169574NB-C22).
\bibliography{references}
\end{document}